\documentclass{SciPost}
\usepackage{hyperref}
\usepackage{graphicx}
\usepackage{lineno}
\usepackage{amsmath}
\usepackage{amssymb}
\usepackage{color}
\usepackage{ulem}
\usepackage[frak=euler,scr=boondox,bb= pazo]{mathalfa}

\def\be{\begin{equation}}
\def\ee{\end{equation}}
\def\bea{\begin{eqnarray}}
\def\eea{\end{eqnarray}}

\begin{document}

\begin{center}{\Large \textbf{%
    Self-organized criticality in neural networks
}}\end{center}

\begin{center}
M. I. Katsnelson\textsuperscript{1}, % m.katsnelson@science.ru.nl
V. Vanchurin\textsuperscript{2,3}, % vitaly.vanchurin@gmail.com
T. Westerhout\textsuperscript{1*}, % tom.westerhout@ru.nl
\end{center}

\begin{center}
{\bf 1} Institute for Molecules and Materials, Radboud University, Heyendaalseweg 135, NL-6525 AJ Nijmegen, The Netherlands
\\
{\bf 2} National Center for Biotechnology Information, NIH, Bethesda, Maryland 20894, USA
\\
{\bf 3} Duluth Institute for Advanced Study, Duluth, Minnesota, 55804, USA
\\
* tom.westerhout@ru.nl
\end{center}

\begin{center}
\today
\end{center}

% For convenience during refereeing: line numbers
%\linenumbers

\section*{Abstract}
% TODO: write your abstract here.
% The abstract is in boldface, and should fit in 8 lines.
% It should be written in a clear and accessible style, emphasizing the context, the problem(s) studied, the methods used, the results obtained, the conclusions reached, and the outlook. You can add a table contents, recommended if your paper is more than 6 pages long.
{\bf
We demonstrate, both analytically and numerically, that learning dynamics of neural networks is generically attracted towards a self-organized critical state. The effect can be modeled with quartic interactions between non-trainable variables (e.g. states of neurons) and trainable variables (e.g. weight matrix). Non-trainable variables are rapidly driven towards stochastic equilibrium and trainable variables are slowly driven towards learning equilibrium described by a scale-invariant distribution on a wide range of scales. Our results suggest that the scale invariance observed in many physical and biological systems might be due to some kind of learning dynamics and support the claim that the universe might be a neural network.}

\vspace{10pt}
\noindent\rule{\textwidth}{1pt}
\tableofcontents\thispagestyle{fancy}
\noindent\rule{\textwidth}{1pt}
\vspace{10pt}

\section{Introduction}\label{sec:introduction}

Many astrophysical \cite{physics}, geological \cite{geology}, biological \cite{biology} and even neurobiological \cite{neurobiology} systems exhibit a remarkable property of being dynamically attracted towards self-organized critical states described by scale-invariant distributions of fluctuations. The self-organized criticality is usually attributed to slowly driven non-equilibrium dynamics of the systems with ``fragile'' equilibrium such as the sand-pile model \cite{BTW,BS,MPB,MZ,BP}, and it is widely believed that the phenomenon is responsible for the emergence of complexity in nature \cite{Bak_book}, including biological complexity \cite{WKK}. In this paper we argue that self-organized criticality is actually an equilibrium phenomenon, but in systems which undergo a learning evolution such as artificial neural networks. A distinctive feature of learning systems is the existence of two different types of degrees of freedom: non-trainable variables that are rapidly driven towards thermodynamic equilibrium and trainable variables that are slowly driven towards learning equilibrium. As we shall see, the learning equilibrium is described by frustrated dynamics on the smallest scales of fluctuations, Gaussian distributions on the largest scales, and scale-invariant distributions on intermediate scales. In what follows, we focus on the analysis of the intermediate scales and the reader is referred to Refs. \cite{theory,world,quantumness} for discussions of other regimes.

Technically, we use a recently developed approach to machine learning from the point of view of statistical physics \cite{theory,world,quantumness} which describes learning as a competition between a general tendency to the entropy growth in physical systems and entropy decrease due to the decrease of information uncertainty during the learning process. Near local equilibrium, the state of the learning system can be described by small fluctuations (both Gaussian and scale-invariant) near the extremal (saddle-point) state determined by this balance, with a dominant contribution of soft modes as is well known in quantum field theory \cite{coleman,shifman,instanton}. We show that it is the soft modes of probability distribution on intermediate scales that are responsible for the scale-invariant fluctuations and the self-organized criticality in neural networks. Keeping in mind deep formal relations between theory of machine learning and fundamental physical theories \cite{theory, world,quantumness} our approach probably can explain a broad distribution of self-organized critical states in completely different systems.

The paper is organized as follows. In Sec. \ref{sec:idea} we describe a mechanism which might be responsible for the emergence of self-organized criticality in the context of a simple model. In Sec. \ref{sec:neural} we argue that the most essential feature of the model is also present in more general learning systems such as artificial neural networks. In Sec. \ref{sec:learning} we show that the learning evolution of neural networks would generically lead to the emergence of self-organized critical states described by a scale-invariant distribution over trainable variables.
%In Sec. \ref{sec:quantum} we consider a near equilibrium dynamics of the trainable variables in the context of emergent quantumness. 
In Sec. \ref{sec:numerics} the main results are verified numerically by following the learning dynamics of artificial neural networks with feedforward architecture. In Sec. \ref{sec:discussion} we discuss implications of the results for machine learning, physics and biology.

\section{Basic mechanism}\label{sec:idea}

In this section we describe the basic mechanism of the self-organized criticality in the context of a simple model with one non-trainable variable $x$ and one trainable variable $q$, but, as we shall see, essentially the same mechanism is responsible for the critical behavior of neural networks with many non-trainable and trainable variables. The main objective of a learning system is to minimize some suitably defined loss function $H(x,q)$ which can be expanded around local extremum up to the second order, i.e.
\be
H(x,q) \approx   \frac{ \lambda_x (x-\bar{x})^2}{2}+ \frac{ g (q-\bar{q})^2 (x-\bar{x})^2}{2} + \frac{ \lambda_q (q-\bar{q})^2}{2} + ... \label{eq:loss0}
\ee
where $\bar x$ and $\bar q$ are mean values of $x$ and $q$ respectively, and $\lambda_x$, $g$ and $\lambda_q$ specify the coupling constants. It is explicitly assumed that there is a quartic coupling between trainable and non-trainable variable, but higher order terms might also be present. We also assume that non-trainable variable $x$ undergoes stochastic dynamics (which maximizes the entropy due to the second law of thermodynamics) and trainable variable $q$ undergoes learning dynamics (which minimizes the entropy due to the second law of learning) \cite{theory}. In the equilibrium the overall entropy remains constant, but the loss function (averaged over the non-trainable variable) depends on the trainable variable:
\be
\int dx \; p(x|q)  H(x,q) = U(q).
\ee
Then, according to the maximum entropy principle, conditional distribution $p(x|q)$ over non-trainable stochastic variable $x$ is given by
\be
p(x|q) \propto \exp \left (- \frac{\beta \lambda_x (x-\bar{x})^2}{2} - \frac{\beta g (q-\bar{q})^2 (x-\bar{x})^2}{2} - \frac{ \beta \lambda_q (q-\bar{q})^2}{2} + ...
 \right )\label{eq:cond}
\ee
and the corresponding free energy is
\bea
F(\beta, q) \approx \frac{1}{2 \beta}\log ( g (q - \bar{q})^2 + \lambda_x )   + \frac{\lambda_q (q-\bar{q})^2}{2} + \frac{1}{2 \beta}\log ( \beta ). \label{eq:free0}
\eea

The next step is to derive marginal distribution $p(q)$, and consequently the joint distribution $p(x,q) = p(x|q) p(q)$. If we assume that both $g$ and $\lambda_q$ are positive, but $\lambda_x$ is negative, then the local minimum of the loss function is at $q=\bar{q}$, but the maximum entropy of $p(x|q)$ is at $q = \bar{q} \pm \sqrt{-\lambda_x/g}$, or when the conditional distribution $p(x|q)$ is nearly flat. In the limit of small fluctuations, $|q-\bar{q}| \lesssim \sqrt{-\lambda_x/g}$, the dynamics of $q$ becomes frustrated: the learning dynamics pushes $q$ towards $\bar{q}$, but the stochastic dynamics pushes $q$ towards either $\bar{q} + \sqrt{-\lambda_x/g}$ or  $\bar{q} - \sqrt{-\lambda_x/g}$. This is exactly the limit when the argument of the logarithm in \eqref{eq:free0} is negative, the conditional distribution $p(x|q)$ has a local maximum at $x=\bar{x}$ and higher order terms must be added to the expansion \eqref{eq:loss0}. Note that the frustration only happens for very small values of $(q-\bar{q})^2 \lesssim -\lambda_x/g $ (or equivalently for small changes of the loss function, $H(q) \propto (q-\bar{q})^2$) where the quantum behavior can emerge (see Ref. \cite{quantumness}), and self-organized criticality is expected to emerge on larger scales $(q-\bar{q})^2 > -\lambda_x/g$ (see Eq. \eqref{eq:dist3}).

According to the first law of learning \cite{theory}, the marginal distribution $p(q)$ must evolve towards a state described by a saddle point of the free energy. If we impose a constraint on the average free energy, 
\be
\int d{q} \; p(q) {F}(\beta, q) = V,\label{eq:constraint2}
\ee
then the equilibrium distribution is 
\be
p(q) \propto e^{- \alpha F({\beta}, q) } \propto ( g (q - \bar{q})^2 + \lambda_x)^{- \frac{\alpha}{2 {\beta}}} \exp\left (-\frac{\alpha \lambda_q (q - \bar{q})^2}{2} \right ).\label{eq:marg2}
\ee
where $\alpha$ is a Lagrange multiplier associated with the constraint \eqref{eq:constraint2}. Evidently, for $\alpha>0$, smaller fluctuations (e.g. on shorter time-scales) would be described by a power-law (or scale-invariant) distribution and larger fluctuations (e.g. on longer time-scales) would be described by a Gaussian distribution, i.e.
\bea
p(q) \propto \begin{cases} ( g (q - \bar{q})^2 + \lambda_x)^{- \frac{\alpha}{2 {\beta}}}  &\;\;\;\text{for} \;\; -\frac{\lambda_x}{g}  < (q-\bar{q})^2 <  \frac{W(\beta \lambda_q)}{\beta \lambda_q}\,,  \\
\exp\left (-\frac{\alpha \lambda_q (q - \bar{q})^2}{2} \right ) &\;\;\;\text{for} \;\;  (q-\bar{q})^2 >  \frac{W(\beta \lambda_q)}{\beta \lambda_q} \,,
\end{cases}\label{eq:dist3}
\eea
where $W(x)$ is the Lambert W function \cite{LambertW}. This suggests that the presence of scale-invariant trainable variables, or, in other words, self-organized critical states, is a direct consequence of the learning dynamics on the intermediate scales, $-\lambda_x/g  < (q-\bar{q})^2 <  W(\beta \lambda_q)/(\beta \lambda_q)$.  

\section{Neural networks}\label{sec:neural}

In the previous section we described a simple model with a scale-invariant distribution over trainable variables, but the learning dynamics which might lead to such a distribution was not yet specified. The key observation was that for the scale-invariance to emerge, interactions between trainable and non-trainable variables must be quartic \eqref{eq:loss0}. It turns out that such interactions are very typical in the context of neural networks for non-trainable states of neurons ${\bf x}$ and trainable elements of the so-called weight matrix $\hat{w}$ (see Eqs. \eqref{eq:bulk_loss} and \eqref{eq:boundry_loss}). Because of the quartic couplings, trainable elements of the weight matrix and bias vector are expected to evolve towards a scale-invariant distribution or, in other words, towards a self-organized critical state.

For numerical tests we will be mainly interested in a feedforward neural architecture (see Sec. \ref{sec:numerics}), but according to analytical results (see Sec. \ref{sec:learning}) it is expected that the same phenomenon would occur in an arbitrary learning system. In general, a neural network can be defined as a septuple $({\bf x}, \hat{P}, p_\partial, \hat{w}, {\bf b}, {\bf f}, H)$, where:
\begin{enumerate}
\item ${\bf x}$, is a (column) state vector of all (input, output and hidden) neurons,  
\item $\hat{P}$, is the boundary projection operators to subspace spanned by input/output neurons,
\item $p_\partial({\bf x}_\partial)$, is a probability distribution which describes the training dataset, 
\item $\hat{w}$, is a weight matrix which describes connections between neurons, 
\item ${\bf b}$, is a (column) bias vector which describes bias in inputs of individual neurons, 
\item ${\bf f}({\bf y})$, is an activation map which describes a non-linear part of the dynamics,
\item ${H}({\bf x}, {\bf b}, \hat{w})$, is a loss function which describes the learning objective.
\end{enumerate} 
We shall refer to all input and output neurons, described by the state vector $\hat{P} {\bf x}$, as boundary neurons, and to all neurons in the hidden layers,  described by the state vector $ (\hat{I} -  \hat{P}) {\bf x}$, as bulk neurons. These different types of neurons evolve according to two different laws \eqref{eq:boundary_eom} and \eqref{eq:bulk_eom}. The state of the boundary neurons depends only on the boundary data, 
\bea
 \hat{P}{\bf x}(t) = \hat{P} {\bf x}_\partial(t), \label{eq:boundary_eom}
\eea
where ${\bf x}_\partial(t)$ is updated either periodically or randomly from a training dataset which can be described by some probability distribution $p_\partial({\bf x}_\partial)$. In contrast, the bulk neurons evolve according to 
\be
(\hat{I} - \hat{P}){\bf x}({t}) = (\hat{I} - \hat{P}) {\bf f} \left(\hat{w} {\bf x}(t-1)+ {\bf b} \right),\label{eq:bulk_eom}
\ee
where the activation map acts separately on each component, i.e. $f_i ( {\bf y} ) = f_i(y_i) $. These functions are called activation functions (e.g. hyperbolic tangent $\tanh(y)$, rectifier linear unit function $\max(0,x)$, etc.) and do not need to be the same for all neurons. 

The main problem in machine learning is to find a bias vector ${\bf b}$ and a weight matrix $\hat{w}$ which minimize (the time-$t$ average or ensemble average over boundary conditions $p_\partial({\bf x}_\partial)$ of) some suitably defined quantity known as the loss function. For example, a boundary loss function could be defined as
\bea
H_\partial({\bf x}, {\bf b}, \hat{w}) &=&  \frac{1}{2} \left ( {\bf x}  - {\bf f} \left ( \hat{w} {\bf x}+ {\bf b} \right) \right )^\dagger  \hat{P} \left (  {\bf x}  - {\bf f} \left ( \hat{w} {\bf x}+ {\bf b} \right) \right ) 
\label{eq:boundry_loss} 
\eea
where because of the inserted projection operator $\hat{P}$, the sum is taken over squared errors at only boundary neurons \cite{theory}. Note that in a feedforward neural architecture there are no errors associated with input neurons and all of the loss comes from the output neurons. Another example is the bulk loss function, e.g.
\bea
H({\bf x}, {\bf b}, \hat{w}) &=&  \frac{1}{2} \left ( {\bf x}  - {\bf f} \left ( \hat{w} {\bf x}+ {\bf b} \right) \right )^\dagger \left (  {\bf x}  - {\bf f} \left ( \hat{w} {\bf x}+ {\bf b} \right) \right ) + \frac{1}{2} {V}({\bf x}, {\bf b}, \hat{w}) 
\label{eq:bulk_loss} 
\eea
where in addition to the first term, which represents a sum of local errors over all neurons, there may be a second term which represents either local objectives or constraints imposed by a neural architecture \cite{theory}. Note that boundary loss is usually used in supervised learning, but bulk loss functions may be used for both supervised and unsupervised learning tasks.

\section{Local equilibrium}\label{sec:learning}

To study the learning dynamics of neural networks analytically it is convenient to switch to a more``macroscopic'' description. Instead of following the individual states we shall study the dynamics of a joint distribution $p({\bf x}, {\bf q})=p({\bf x}| {\bf q}) p({\bf q}) $ over non-trainable variables ${\bf x}$, which describe the current state vector of $N$ neurons, and trainable variables ${\bf q}$, which describe the state of $K$ dynamical elements of weight matrix ${\bf w}({\bf q})$ and bias vector ${\bf b}({\bf q})$. If we fix the trainable variables ${\bf q}$ and impose a constraint on the average loss function 
\be
\int d^N x H({\bf x},{\bf q}) p({\bf x}|{\bf q}) = U({\bf q}) \label{eq:U},
\ee
then the maximum entropy probability distribution is given by
\be
p({\bf x}|{\bf q}) \propto \exp \left (- \beta  H({\bf x},{\bf q}) \right ). \label{eq:dist0}
\ee
The bulk loss function \eqref{eq:bulk_loss} can be expanded around a local extremum $(\bar{\mathbf{x}}, \bar{\mathbf{q}})$ as
\be
H({\bf x},{\bf q}) =\frac{1}{2} ({\bf x} - \bar{\bf x})^\dagger \left (\hat{G}({\bf q}) + \hat{V}_x \right ) ({\bf x} - \bar{\bf x}) +\frac{1}{2}  ({\bf q} - \bar{\bf q})^\dagger \hat{V}_q ({\bf q} - \bar{\bf q}),
\ee
where
\be
\hat{G}({\bf q})  \equiv \left ( \hat{I} - \hat{F} \hat{w}\right)^\dagger \left ( \hat{I} - \hat{F} \hat{w}\right).
\ee
For simplicity we assume that $\hat{F}$, $\hat{V}_x$ and $\hat{V}_q$ are diagonal matrices of, respectively, first derivatives of activation functions $f_i(y_i)$ and second derivatives of local potentials $V({\bf x},{\bf q})$ with respect to non-trainable ${\bf x}$ and trainable ${\bf q}$ variables. Then the free energy of the maximum entropy distribution \eqref{eq:dist0} is given by \cite{theory},
\be
 {F}({\beta}, {\bf q}) = \frac{1}{2\beta} \log \det \left(\beta \hat{G} ({\bf q}) + \beta \hat{V}_x  \right) +  \frac{1}{2} ({\bf q} - \bar{\bf q})^\dagger \hat{V}_q ({\bf q} - \bar{\bf q}) \label{eq:lfree}.
\ee

The next step is to determine the marginal distribution $p({\bf q})$ and consequently the joint distribution $p({\bf x},{\bf q})=p({\bf x}|{\bf q}) p({\bf q})$. If we impose a constraint on the free energy
\be
\int d{\bf q} \; p({\bf q}) {F}({\beta}, {\bf q})  = V,
\ee
then the marginal distribution is 
\be
p({\bf q}) \propto \exp\left (- \alpha  {F}({\beta}, {\bf q})  \right ).\label{eq:dist}
\ee
For free energy \eqref{eq:lfree} the distribution is 
\be
p({\bf q}) \propto  \left[ \det \left( \hat{G} ({\bf q}) + \hat{V}_x  \right) \right]^{-\frac{\alpha}{2\beta}} \exp\left (  -  \frac{\alpha}{2} ({\bf q} - \bar{\bf q})^\dagger \hat{V}_q ({\bf q} - \bar{\bf q})   \right ).
\ee 
Since in the learning equilibrium the free energy would be extremized, it makes sense to expand individual eigenvalues $\lambda_i({\bf q})$ of the matrix $\hat{G}({\bf q})+\hat{V}_x$ around local extrema, 
\be
\lambda^i({\bf q})  \approx \lambda^i_0 + \left ( {\bf q} - \bar{\bf q}\right )^\dagger \hat{\lambda}^i  \left ( {\bf q} - \bar{\bf q}\right ),\label{eq:stochastic}
\ee
and then the distribution \eqref{eq:dist} can be approximated as
\be
p({\bf q}) \propto   \prod_i \left ( \lambda^i_0 + \left ( {\bf q} - \bar{\bf q}\right )^\dagger \hat{\lambda}^i  \left ( {\bf q} - \bar{\bf q}\right ) \right  )^{-\frac{\alpha}{2\beta}} \exp\left (  -  \frac{\alpha}{2} ({\bf q} - \bar{\bf q})^\dagger \hat{V}_q ({\bf q} - \bar{\bf q})   \right ).\label{eq:three-regimes}
\ee
Evidently, the marginal distribution $p({\bf q})$ would be scale-invariant (or a power-law) on intermediate scales when the first term dominates and Gaussian on larger scales when the second term dominates. Therefore, the scale-invariant and Gaussian distributions are expected for, respectively, short-term and long-term dynamics. Also note that for very small fluctuations (e.g. when $\left ( {\bf q} - \bar{\bf q}\right )^\dagger \hat{\lambda}^i  \left ( {\bf q} - \bar{\bf q}\right )$ and $\lambda^i_0$ are of the same order but have opposite signs) the free energy \eqref{eq:lfree} might diverge and the system can become frustrated from simultaneous maximization of the entropy of non-trainable variables and minimization of the entropy of trainable variables. (See Sec. \ref{sec:idea} for a discussion of this point in the context of a simple model.) %In the following section we shall analyze such systems in the context of emergent quantumness.  

\section{Numerical results}\label{sec:numerics}

In this section we will justify the model from Sec. \ref{sec:learning} by training
a feedforward neural network until it reaches equilibrium and then analyzing its
behavior in equilibrium.

We focus on the MNIST dataset \cite{MNIST} which is a collection of images of
handwritten digits. It contains 60000 training and 10000 test samples. We use a
simple fully-connected feedforward neural network with two hidden layers with 700
and 476 neurons and ReLU \cite{ReLU} activation function. This network is optimized
using stochastic gradient descend with batch size $1$ and learning rate $2 \cdot
10^{-3}$. We also add $L_2$-regularization term to the loss function with rate
$5 \cdot 10^{-4}$. The training proceeds for $\mathcal{O}(3000)$ epochs (i.e.
passes through the dataset) to reach equilibrium. We keep track of the loss
function on the test dataset and ensure that no overfitting takes place.

\begin{figure}[h!]
    \centering
    \includegraphics[width=12cm]{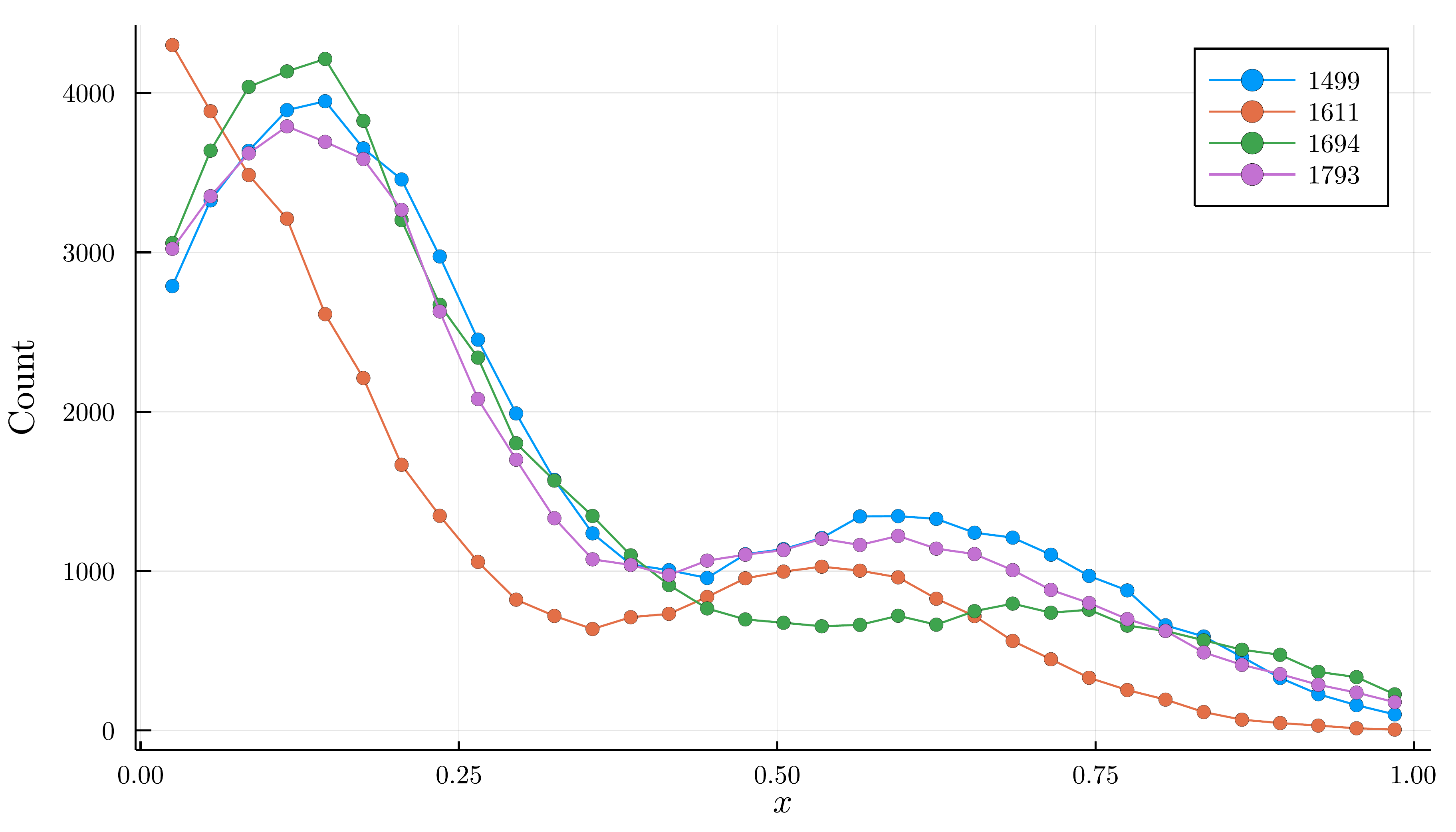}
    \caption{%
        Output distribution $p(x|\mathbf{q})$ for a few randomly selected neurons
        with frozen weights (i.e. no training takes place). All neurons come from
        the second hidden layer.
    }
    \label{fig:neuron-output-distribution}
\end{figure}

\begin{figure}[h!]
    \centering
    \includegraphics[width=12cm]{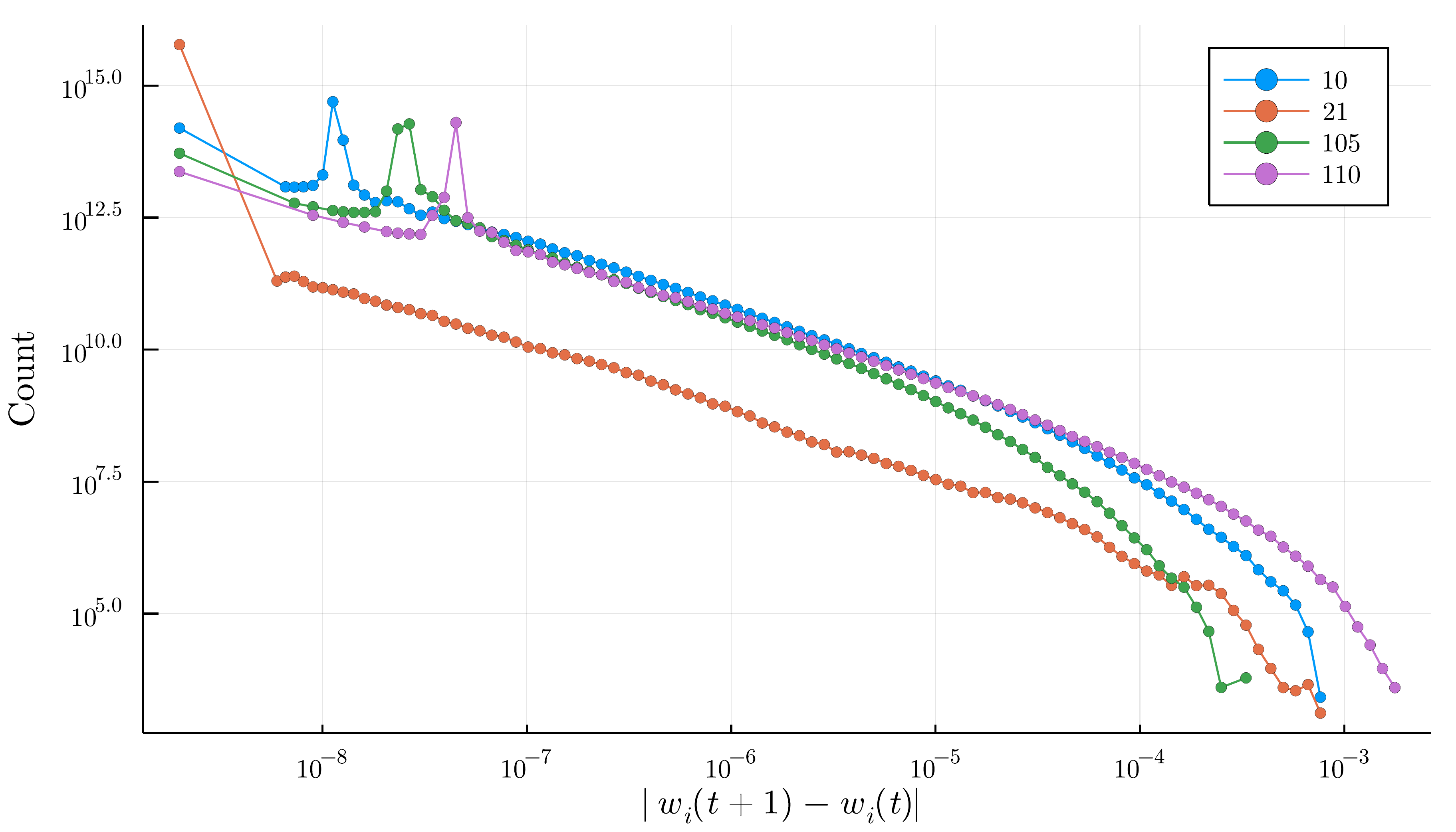}
    \caption{%
        Distribution of local fluctuations for a few randomly selected weights from
        the first and second layer.
    }
    \label{fig:momentum-distribution}
\end{figure}

We can now directly compute $p(x|\mathbf{q})$ by freezing the neural network
parameters and tracking neuron outputs for different input images. The result
is shown in Figure \ref{fig:neuron-output-distribution}. We clearly see two
maxima which correspond to the frustrated regime discussed in Sec. \ref{sec:learning}.
This regime is characterized by negative $\lambda^i_0$ from \eqref{eq:stochastic} which,
because of \eqref{eq:dist0}, appear as local minima in $p(x|\mathbf{q})$. 

Next, we train the neural network for another 200 epochs and keep track of a few
randomly selected weights. As a result we have time dependence $w_i(t)$ of trainable
parameters in equilibrium. This data can be used to gain insight into $p(\mathbf{q})$.

\begin{figure}[h]
    \centering
    \includegraphics[width=15cm]{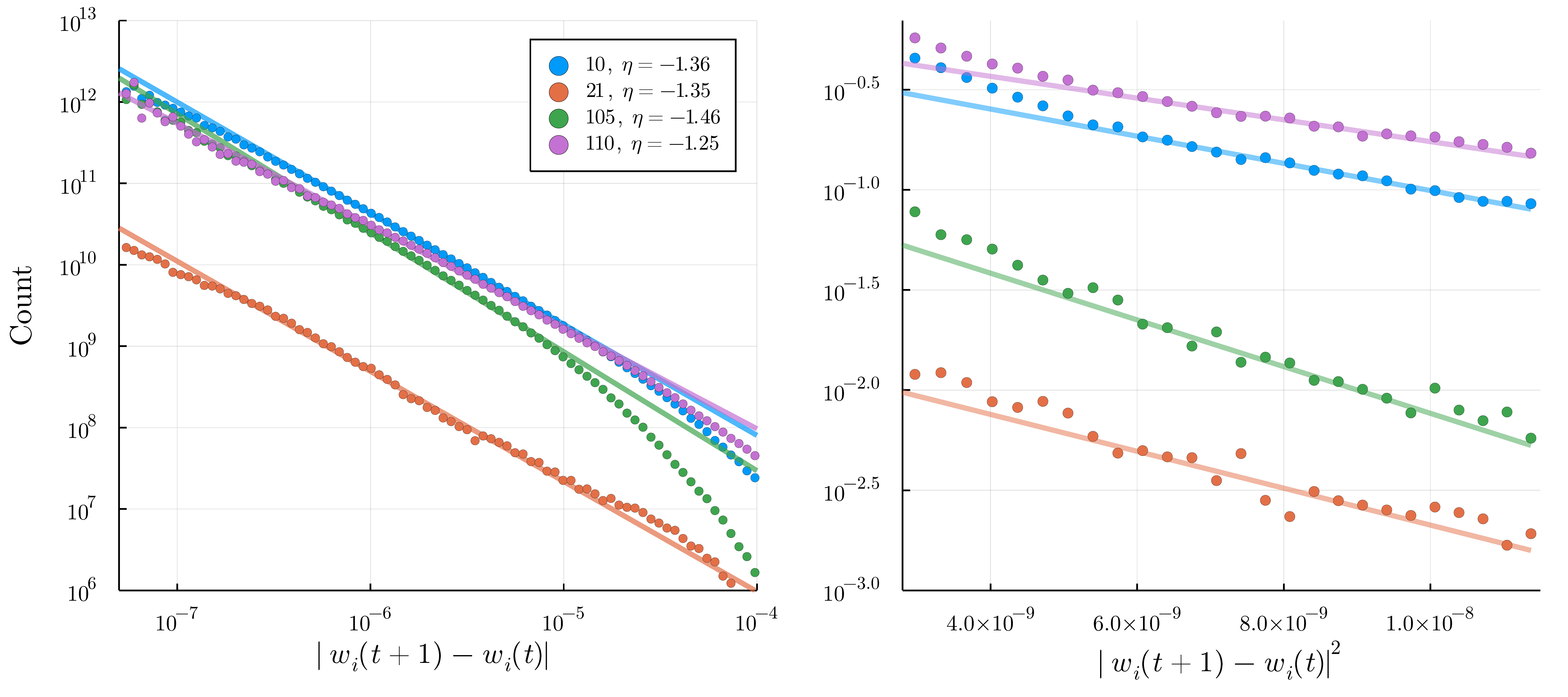}
    \caption{%
        Distribution of local fluctuations for a few randomly selected weights. This figure contains the exact same data as Figure \ref{fig:momentum-distribution}, but focuses
        on intermediate (left panel) and large (right panel) fluctuations. Left panel shows data in log-log scale and straight lines correspond to power-law behavior. Right panel shows data in log-quadratic scale and straight lines correspond to Gaussian decays.
    }
    \label{fig:momentum-distribution-zoom}
\end{figure}

In Sec. \ref{sec:learning} we expanded everything around $\mathbf{\bar q}$. In
reality, the situation might be more complicated because there might be many
extrema close to one another such that in equilibrium the neural network constantly
``hops'' between them. To account for this hopping we will consider truly local
fluctuations: $w_i(t+1) - \bar w_i(t) \approx w_i(t+1) - w_i(t)$. The distribution
of these fluctuations is shown in Figure \ref{fig:momentum-distribution}. We see a
peak for very small fluctuations, power-law decay at intermediate values, and
Gaussian decay for large fluctuations. In Figure \ref{fig:momentum-distribution-zoom}
we focus on the intermediate and large fluctuations to better illustrate the
power-law and Gaussian decays. This behavior matches \eqref{eq:three-regimes}
perfectly.

\begin{figure}[h!]
    \centering
    \includegraphics[width=12cm]{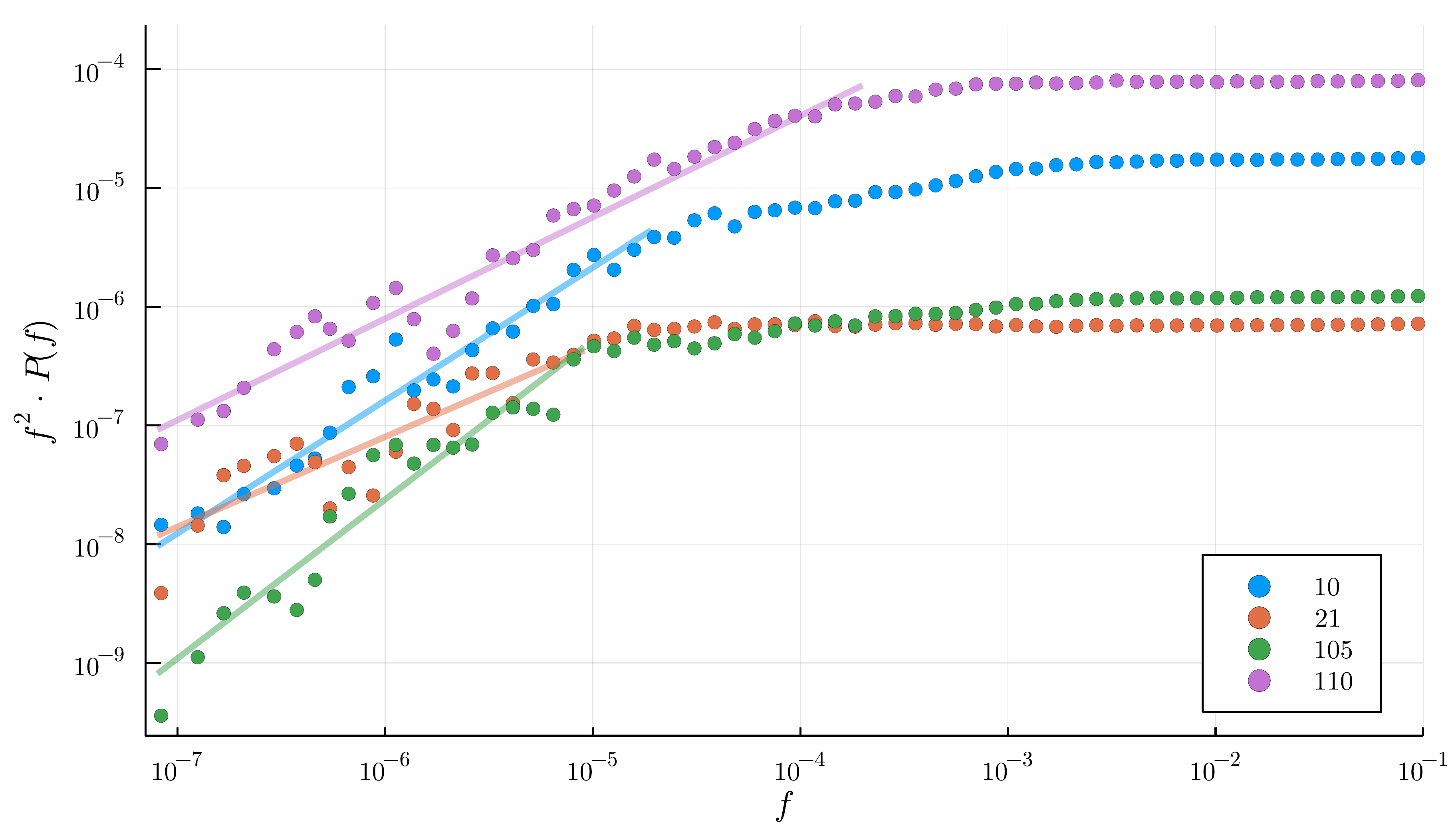}
    \caption{%
        Power spectrum of $w_i(t)$. We show $f^2 \cdot P(f)$ rather than
        $P(f)$ to more clearly indicate the deviations from $1/f^2$ behavior
        which is expected for completely uncorrelated jumps. Straight
        lines are guide to eye and indicate regions where the power spectrum
        decays as $1/f^a$ with $a < 2$.
    }
    \label{fig:fourier-power-spectrum}
\end{figure}

Finally, we analyze the power spectrum of $w_i(t)$, i.e.
\bea
P(f) &\equiv&  \left |\int_0^T dt  w_i(t) e^{-i f t}\right|^2  = 
 \frac{1}{f^2}  \left | \int_0^T dt\;  \frac{d w_i(t)}{dt}e^{i f t} \right |^2 \,.
\eea
%\bea 
%P(f) &\equiv& \left |\int_0^T dt  w_i(t) e^{-i f t}\right|^2 = \int_0^T dt w_i(t)e^{i f t} \int_0^T dz w_i(z) e^{-i f z} \notag \\
%&=& \int_0^T dt \; w_i(t)  \frac{1}{i f} \frac{d}{dt} e^{i f t}  \int_0^T dz\;  w_i(z)  \frac{1}{-i f} \frac{d}{dz}e^{- i f z} \notag \\
%&=& \frac{1}{f^2} \left ( \int_0^T dt\;  \frac{d w_i(t)}{dt}e^{i f t} \int_0^T dz\;  \frac{d w_i(z)}{dz}e^{-i f z} \right )\notag \\
%&=& \frac{1}{f^2} \left | \int_0^T dt\;  \frac{d w_i(t)}{dt}e^{i f t} \right |^2,
%\eea 
$\left \langle \left | \int_0^T dt\;  \frac{d w_i(t)}{dt}e^{i f t} \right |^2 \right \rangle$ is constant for independent increments, i.e. when
\be
\left \langle \frac{d w_i(t)}{dt} \frac{d w_i(z)}{dz} \right \rangle \propto \delta(t-z) \label{eq:uncorrelated}
\ee
as would be the case, for example, for white noise. In Figure \ref{fig:fourier-power-spectrum} we show
$f^2 \cdot P(f)$ as a function of $f$ such that $1/f^2$ dependence would correspond to a horizontal line. On short time scales (corresponding to large $f$) the system is in a local minimum and the dynamics of $w_i(t)$ can be modeled with uncorrelated jumps described by power spectrum $P(f)\propto 1/f^2$. On the long time scales (small $f$) the system is hopping between local minima and the jumps are are described by power spectrum $P(f)\propto 1/f^a$ with $a<2$. (See Refs. \cite{theory, world, quantumness} for analytical modeling of the learning dynamics of $p(q,t)$ and $F(q,t)$ on both short and long time-scales.)

\section{Discussion}\label{sec:discussion}

In this article, we analyzed, both numerically and analytically, the learning dynamics of neural networks near equilibrium and showed that the learning systems are generally attracted towards critical states described by scale-invariant distributions over trainable variables on a wide range of scales. Moreover, on even larger scales the trainable degrees of freedom behave as Gaussian random variables and on somewhat smaller scales the dynamics is frustrated from simultaneous maximization of the entropy of non-trainable variables and minimization of the entropy of trainable variables. These results have some interesting implications for machine learning, physics and biology.  

{\it Machine learning.} Every trainable variable evolves towards a state described by a scale-invariant distribution on a range of scales, but the range itself can be very different for different variables. Some variables are very well trained and have a scale-invariant distribution on a wide range of scales, whereas others are poorly trained and the scale-invariant range is very narrow. This suggests that the size of the range can be used to determine how well a given trainable variable was trained or how vital it is for the overall performance of the network. For example, if the least vital trainable variables (i.e. with the smallest ranges of scale-invariant distributions) can be successfully identified, then the learning algorithm can be improved or the neural network can be compressed by either rewiring or dropping the least vital connections. On a more practical level, the scale-invariant range is usually smaller for trainable variables whose amplitude of fluctuations is larger (and vise versa), and therefore the amplitude can be used to identify the least vital connections that should be either dropped (for compressing neural networks) or rewired (for improving efficiency of learning). 

{\it  Physics.} It was recently proposed that the entire universe may be a neural network which undergoes a learning evolution \cite{world}. If correct, then all of the physical phenomena are not fundamental, but rather emergent, and provide an adequate description in the limit of large number of degrees of freedom, e.g. neurons, weights, biases etc. In particular, it was shown that the learning dynamics near equilibrium can be modeled using either thermodynamics \cite{theory} or quantum mechanics \cite{quantumness} and further away from the equilibrium using either Hamiltonian mechanics or general relativity \cite{world}. In this paper we uncovered yet another near-equilibrium limit in which a self-organized criticality emerges on intermediate scales from the learning dynamics of neural networks. Given that the self-organized criticality is observed in many physical (and also biological) systems, our results support the claim that the entire universe may be a neural network. 

{\it Biology.} Self-organized criticality in the context of biological complexity was recently discussed in Ref. \cite{WKK}. It was suggested that competing interactions between different levels of organization form a universal mechanism leading to biological complexity. At the same time, mathematical mechanisms responsible for the self-organized criticality in biology \cite{BS,BP,Bak_book} are still very poorly studied. It is a big temptation to identify, in some sense, Darwin selection with learning and apply the theory of machine learning to evolutionary biology. This exciting issue is far beyond the scope of this work and will be considered elsewhere. 
\\

% \section*{Acknowledgements}

% TODO: include author contributions
% \paragraph{Author contributions}
% This is optional. If desired, contributions should be succinctly described in a single short paragraph, using author initials.

% TODO: include funding information
% Authors are required to provide funding information, including relevant agencies and grant numbers with linked author's initials. Correctly-provided data will be linked to funders listed in the \href{https://www.crossref.org/services/funder-registry/}{\sf Fundref registry}.
\paragraph{Funding information}
V.V. was supported in part by the Foundational Questions Institute (FQXi) and the Oak Ridge Institute for Science and Education (ORISE). The work of M.I.K. and T.W. was supported by European Research Council via Synergy Grant 854843 - FASTCORR. Numerical simulations in this work were carried out on the Dutch national \mbox{e-infrastructure} with the support of SURF Cooperative.

\bibliography{references.bib}

\nolinenumbers

\end{document}